\documentclass[12pt,english,dvips]{scrartcl}
\usepackage{babel}
\usepackage[T1]{fontenc}
\usepackage[ansinew]{inputenc}

\begin{document}

\begin{titlepage} \vspace{0.2in} 

\begin{center} {\LARGE \bf 
Cosmological Issues for Revised 
Canonical Quantum Gravity \\} \vspace*{0.8cm}
{\bf Giovanni Montani}\\ \vspace*{1cm}
ICRA---International Center for Relativistic Astrophysics\\ 
Dipartimento di Fisica (G9),\\ 
Universit\`a  di Roma, ``La Sapienza",\\ 
Piazzale Aldo Moro 5, 00185 Rome, Italy.\\ 
e-mail: montani@icra.it\\ 
\vspace*{1.8cm}

PACS 83C  \vspace*{1cm} \\ 

{\bf   Abstract  \\ } \end{center} \indent
In a recent work \cite{M02} we presented a reformulation
of the canonical quantum gravity, based on adding the so-called
{\em kinematical term} to the gravity-matter action;
this revised approach leads to a self-consistent canonical
quantization of the 3-geometries, referred to the
external time as provided via the added term.

Here, we show how the {\em kinematical term}
can be interpreted in terms
of a non relativistic dust fluid which plies the role
of a ``real clock' for the quantum gravity theory, and,
in the WKB limit of a cosmological problem, makes
account for a {\em dark matter} component which, at present
time, could play a dynamical role. 

\end{titlepage}

\section{Fundamental Remarks}

The Canonical Method of quantization relies on
the existence of an Hamiltonian function for the system,
regarded as conjugate variable to the physical time,
and is implemented by recognizing the dynamics admits
an Hamiltonian constraint; in fact, the quantum dynamics
is easily obtained by the transcription of such a constraint
via the operators associated to the canonical variables \cite{K81}.

As an helpful example for the analysis below developed,
we review the case of the one-dimensional non-relativistic
(parametrized) particle, whose action reads 

\begin{equation}
S=\int \{ p\dot{q} - h(p, q)\} dt \, ,
\label{a}
\end{equation}

where $t$ denotes the Newton time and $h$ the
Hamiltonian function. In order to quantize this system,
we parameterize the Newton time as $t=t(\tau )$,
so getting the new action as

\begin{equation}
S=\int \{ p\frac{dq}{d\tau } - h(p, q)\frac{dt}{d\tau }\} d\tau
\, .
\label{b}
\end{equation}

Now we set $p_0\equiv -h$ and add this relation to the
above action by a Lagrangian multiplier $\lambda $, i.e.

\begin{equation}
S=\int \{ p\frac{dq}{d\tau } + p_0\frac{dt}{d\tau } -
\bar{h} (p, q, p_0, \lambda )\} d\tau
\, \quad \bar{h} \equiv \lambda (h + p_0)
\, .
\label{c}
\end{equation}

By varying this action with respect to $p$ and $q$,
we get the Hamilton equations
$dq/d\tau = \lambda \partial h/\partial p$ and
$dp/d\tau = -\lambda \partial h/\partial q$, while
the variations of $p_0$ and $t$ yield
$dt/d\tau = \lambda$ and $dp_0/d\tau = 0$;
all together, these equations describe the same Newton
dynamics, having the energy as constant of the motion.
But now, by varying $\lambda$, we get the (desired)
constraint $h + p_0 = 0$, which, in terms of the
operators $\hat{p}_0 = -i\hbar \partial _t$ and $\hat{h}$,
provides the Schr\"odinger equation
$i\hbar \partial _t \psi = \hat{h}\psi$, as taken for the system
state function $\psi (t, q)$.
Finally we remark that, when retaining the relation
$dt/d\tau = \lambda$, we are able to write the
wave equation in the parametric time as

\begin{equation}
i\hbar \partial _{\tau } \psi (\tau , q) = \lambda (\tau ) 
\hat{h}\psi (\tau , q) 
\, ,
\label{d}
\end{equation}

where $\lambda (\tau )$ is to be assigned.\\ 
In spite of its simplicity, this example is a naive, 
but very good prototype of our approach to the canonical
quantum gravity.

When considering the gravitational field, the notion
of an Hamiltonian function is recognized, as soon as,
the four-dimensional manifold ${\cal M}^4$ is splitted into a
one-parameter family of spatial hypersurfaces $\Sigma ^3$, i.e.
${\cal M}^4=\Sigma ^3\otimes R$ \cite{ADM62}.\\
Thus, if ${\cal M}^4$ admits generic internal coordinates
$y^{\rho }$ and a metric tensor $g_{\mu \nu }(y^{\rho })$
($\mu , \nu , \rho = 0,1,2,3$), then we can chose a family of
space-like hypersurfaces filling ${\cal M}^4$, by assigning the
parametric equations $y^{\rho }=y^{\rho }(t, x^i)$
($i=1,2,3$); so doing, we obtain a new basis, composed of the
normal field $n^{\mu }(y^{\rho })$ and the tangent vectors
$e^{\mu }_i\equiv \partial _iy^{\mu }$ to $\Sigma ^3$
($g_{\mu \nu } n^{\mu }n^{\nu } = -1 , \;
g_{\mu \nu } n^{\mu }e^{\nu }_i = 0$), on which, 
we can project the deformation vector
$N^{\mu } \equiv \partial _t y^{\mu }$ as
$\partial _t y^{\mu } = Nn^{\mu } + N^i\partial _iy^{\mu }$
(where $N$ and $N^i$ are respectively called the lapse
function and the shift vector).\\
Now we may regard the parameterization
$y^{\rho }=y^{\rho }(t, x^i)$, as a coordinate transformation
between two coordinates systems and, observing that
the 3-metric induced on $\Sigma ^3$ reads
$h_{ij}\equiv g_{\mu \nu }
\partial _i y^{\mu }\partial _j y^{\nu }$,
finally, the line element admits the representation

\begin{equation}
ds^2=g_{\mu \nu }dx^{\mu }dx^{\nu}=
-N^2dt^2 + h_{ij}(dx^i + N^idt)(dx^j + N^jdt)
\, .
\label{e}
\end{equation}

In the system of coordinates $t$ and $x^i$, the
ten independent $g_{\mu \nu }$ are replaced by the ten
independent functions $\{ N, N^i, h_{ij}\}$, and the normal
field takes the form $n^{\mu }=\{ 1/N, -N^i/N\}$.

When the Einstein-Hilbert and matter
(below we consider a real scalar field)
action is recasted
in this system of coordinates, it happens that the variables
$N$ and $N^i$, being cyclic ones, behave as Lagrange multipliers.
As a consequence, we get eight constraints, corresponding
to the vanishing of the conjugate momenta
$p_N$ and $p_{N^i}$, as well as, of the super-Hamiltonian
$(H^g + H^{\phi })$ and of the super-momentum
$(H^g_i + H^{\phi }_i)$ (the labels $g$ and $\phi$ refer
respectively to the metric and the scalar field).\\
The canonical quantization of the system, is then performed by
upgrading the 3-metric and its conjugate momentum to operators
$\hat{h}_{ij}, \hat{\pi }^{ij}=-i\hbar
\delta(\quad )/\delta h_{ij}$
(the same for the field variables leads to $\hat{\phi }$
and $\hat{p}_{\phi }=-i\delta (\quad )/\delta \phi $), and
implementing the Hamiltonian constraint on a quantum level, i.e.
$(\hat{H}^g + \hat{H}^{\phi })\Psi = 0 \quad 
(\hat{H}^g_i + \hat{H}^{\phi }_i)\Psi = 0$. 
The former of these functional equations, known as 
the Wheeler-DeWitt one, provides the quantum dynamics,
while the latter ensures the invariance under the
3-diffeomorphisms, i.e. the state function $\Psi$ depends
on $\phi $ and a class of 3-geometries $\{ h_{ij}\}$
\cite{D67}
\footnote{The fact that, the wave functional is independent of
$N$ and $N^i$, reflects the classical constraints
$p_N=0$ add $p_{N^i}=0$.}.\\
The most unsatisfactory features of the above
formulation consist of the absence of a time evolution and
the impossibility for an Hilbert space \cite{K81,D97}
(see also \cite{I92} and \cite{R91a,R91b}); but it exists
also a ground-level criticism: indeed, the
above splitting procedure relies on the possibility to
distinguish between space-like and time-like objects
(for instance the normal field $n^{\mu }$ should be time-like),
but, when $g_{\mu \nu}$ is a quantum field, these notions 
can be recognized at most in terms of
``expectation values''. Therefore, on a quantum level, the
(3 + 1)-splitting makes sense only in a perturbative limit, 
when yet survives the concept of metric background. 
A different approach consists of determining
the character of geometrical objects, before
quantizing the system, i.e. fixing the reference frame,
determining the space-like family of hypersurfaces
$\Sigma ^3$ and only then quantizing the dynamics.//
The following three sections answer the fundamental
questions concerning such a different point of view,i.e.: 
what physically it means? Which quantum dynamics it yields?
Which cosmological issues it predicts?

\section{The Kinematical Action and the Reference Fluid}

To fix the reference frame, it is equivalent
to assign the lapse function and the shift vector already in the
action, and then varying only the 3-metric;
but so doing, we loss the super-Hamiltonian and the
super-momentum constraints and, though, we have to do with
$\infty^3$ degrees of freedom, nevertheless our situation
becomes very similar to that one analyzed for
a non-relativistic particle.
Thus, in close analogy to such case, we parametrized the
gravity-matter action by means of the so-called
``kinematical term'' \cite{K81}, which refers the dynamics to
the generic coordinates $y^{\rho }$, playing here the role
that above was proper of the Newton time.
The extended action reads \cite{M02}

\begin{equation}   
S^{g\phi k} = \int _{{\cal M}^4} \left\{ 
\pi ^{ij}\partial _th_{ij}  + \pi _{\phi }   
\partial _t\phi + p_{\mu }\partial _ty^{\mu } -  
N(H^g + H^{\phi } + p_{\mu }n^{\mu }) - N^i(H^g_i + H^{\phi }_i + 
p_{\mu }\partial _iy^{\mu })\right\} d^3xdt  
\, . 
\label{f}
\end{equation}  

Above, the (normal) field $n^{\mu }(y^{\rho })$
is, on this level, to be regarded as a generic one and
therefore assigned arbitrarily; 
$\partial _iy^{\mu }$ is a potential-like term,
$p_{\mu }$ is determined by the field equations and
anyway all the added terms are metric independent.\\ 
By varying this action with respect to $h_{ij}\; {\pi }_{ij}\;
\phi \; \pi _{\phi }$, we get the field equations as unchanged,
while the variation of $N$ and $N^i$ provides the
new (desired) constraints

\begin{equation}   
H^g + H^{\phi } = -p_{\mu }n^{\mu }
\, \quad 
H^g_i + H^{\phi }_i = -p_{\mu }\partial _iy^{\mu }
\label{g}
\end{equation}

Finally, the variation of the kinematical variables
$y^{\mu }$ and $p_{\mu }$, provides the additional equations 

\begin{equation} 
\partial _ty^{\mu } = Nn^{\mu } + 
N^i\partial _iy^{\mu } \, \quad
\partial _tp_{\mu } =
-Np_{\rho }\partial _{\mu }n^{\rho } + 
\partial _i(N^ip_{\mu })\, .
\label{h}
\end{equation} 

Since $n^{\mu }$ is assigned {\em ab initio}, then the
specification of $N$ and $N^i$
allows to solve the first of these equations for
$y^{\mu }(t, x^i)$ and hence, the second one, yields the
generalized momentum $p_{\mu }(t, x^i)$ (entering linearly
in the Hamiltonian).\\

Now, to understand the physical meaning of the added kinematical
term, we have to investigate these field equations by restoring
their covariant form, via the variables $y^{\mu }$.
To this end, we denote the coordinates ($t, x^i$) by barred
indices $\bar{\mu }, \bar{\nu }, \bar{\rho },...$
and we remark that the following relations take place:
$\partial _t = \partial _ty^{\mu }\partial _{\mu }\;
\partial _i = \partial _iy^{\mu }\partial _{\mu }\;
n^{\bar{\mu }}\partial _{\bar{\mu }}=
n^{\mu }\partial _{\mu }$.
Then the first of equations (\ref{h}) rewrites as
$n^{\mu } = n^{\bar{\rho }}\partial _{\bar{\rho }}y^{\mu }$;
this equation is crucial to ensure that, after the variation,
$n^{\mu }$ is a real unit time-like vector, i.e.

\begin{equation}
g_{\mu \nu }n^{\mu }n^{\nu } = 
g_{\mu \nu }n^{\bar{\rho }}\partial _{\bar{\rho }}y^{\mu }
n^{\bar{\sigma }}\partial _{\bar{\sigma }}y^{\nu } = 
g_{\bar{\rho } \bar{\sigma } }n^{\bar{\rho }}
n^{\bar{\sigma }} = -1 \, ,
\label{un}
\end{equation}

the last equality being true by construction of
$g_{\bar{\mu }\bar{\nu }}$ and $n^{\bar{\mu }}$;
in fact the metric form $g_{\bar{\mu }\bar{\nu }}$, as
given in the line element (\ref{e}), ensures the normal vector
$n^{\bar{\mu }} \equiv (1/N, \; -N^i/N)$ be a unit timelike one.\\
Thus, after the variation of the action, our approach ensures,
differently from the Wheeler-DeWitt one, that we have 
to do, even on a quantum level. with a real normal field, 
as far as the first kinematical equation holds. 

Being $n^{\mu }p_{\mu }$ a 3-scalar density of weight $1/2$,
we may set $n^{\mu }p_{\mu }\equiv -\omega (t, x^i)$, 
and then define the real scalar
$\varepsilon \equiv -\omega /\sqrt{h}$
(with $h\equiv deth_{ij}$).\\
Using these information and the first  of (\ref{h}),
we may rewrite the second one as

\begin{equation} 
n^{\rho }[\partial _{\rho }(Np_{\mu }) -
\partial _{\mu }(Np_{\rho })] = 
-\partial _{\mu }(N\sqrt{h}\varepsilon ) + p_{\mu }
(n^{\rho }\partial _{\rho }N + \partial _iN^i)
\,  . 
\label{l}
\end{equation} 

Since the fundamental aim of our reformulation,
apart from removing the ambiguity about the time-like
character of the normal, consists of constructing a quantum
gravity theory having evolution,
there are no serious reasons 
to deform the super-momentum constraint; 
indeed, as we shall see better in the next section,
the ``frozen formalism'' is removed as soon as
the Super-Hamiltonian is no longer zero. 
Therefore we can require the condition
$p_{\mu }\partial _iy^{\mu }\equiv p_i=0$ holds, so
getting $p_{\mu }=\omega n_{\mu }= 
-\sqrt{h}\varepsilon n_{\mu }$;
by other words, such a conservative choice, is equivalent to
take the generalized momentum as parallel to the normal field, and 
therefore time-like. 
The above form of the generalized momentum is preserved
by the dynamics
(i.e., once it is assigned as a cauchy problem,
then the second of the kinematical equations ensures it holds
for ever), 
under the constraint $\partial _iN=0$. In fact, by multiplying
the second of equations (\ref{h}) by $\partial _iy^{\mu }$,
we get, after some algebra, 
the following equation for $p_i$

\begin{equation}
\partial _tp_i - p_j\partial _iN^j - \epsilon \sqrt{h}
\partial _iN = 0 \, .
\label{ppp}
\end{equation} 

We see how the above condition transforms (\ref{ppp})
into a linear and
homogeneous partial differential equation (in normal form)
in the unknowns $p_i$;
if we set the initial condition $p_i^0=0$,
then the unique solution is $p_i\equiv 0$, valid for any
later time.\\ 
Remarking that, in the barred coordinates
$N\sqrt{h}=\sqrt{-\bar{g}}$,
the above restriction
$p_{\mu } = -\sqrt{h}\varepsilon n_{\mu }$ 
reduces equation (\ref{l}), via some technical steps,
to the form

\begin{equation} 
\varepsilon n^{\rho }(\partial _{\rho }n_{\mu } -
\partial _{\mu }n_{\rho }) +
\frac{n_{\mu }}{\sqrt{-g}}\partial _{\rho }(\sqrt{-g}
\varepsilon n^{\rho }) = 0
\,  . 
\label{m}
\end{equation} 

The above equation reads covariantly as 

\begin{equation} 
\varepsilon n^{\rho }(\nabla _{\rho }n_{\mu } - 
\nabla _{\mu }n_{\rho }) +
n_{\mu }\nabla _{\rho }(\varepsilon n^{\rho }) = 0 
\label{n}
\end{equation} 

which, since
$n^{\rho }\nabla _{\mu }n_{\rho }=0$
(being $n^{\mu }$ a unit vector), finally becomes

\begin{equation} 
\varepsilon n^{\rho }\nabla _{\rho }n_{\mu } + 
n_{\mu }\nabla _{\rho }(\varepsilon n^{\rho }) = 0 =
\nabla _{\rho }t^{\rho }_{\mu }
\, . 
\label{o}
\end{equation} 

Thus we got the surprising result that, the
kinematical momentum equation, 
reduces to the conservation law of a ''dust energy-momentum
tensor'' $t_{\mu \nu }=\varepsilon n_{\mu }n_{\nu }$.
Multiplying equation (\ref{o}) by $n^{\mu }$,
it implies the additional condition 

\begin{equation} 
\nabla _{\rho }(\varepsilon n^{\rho }) = 0 
\, ; 
\label{p}
\end{equation} 

this condition simplifies (\ref{o}) to the form 

\begin{equation} 
n^{\rho }\nabla _{\rho }n_{\mu } = 0 =
n^{\rho }(\partial _{\rho }n_{\mu } - \frac{1}{2}
n^{\sigma }\partial _{\mu }g_{\rho \sigma }) 
\, . 
\label{q}
\end{equation} 

As well known, the dynamical equations describing
a perfect fluid
characterized by energy density $\rho$, pressure $p$, entropy
density $\sigma $ and four-velocity $u_{\mu }$, take the form

\begin{equation} 
\nabla _{\mu }(\sigma u^{\mu }) = 0 
\label{r}
\end{equation} 

\begin{equation} 
(\rho + p)
u^{\rho }(\partial _{\rho }u_{\mu } - \frac{1}{2}
u^{\sigma }\partial _{\mu }g_{\rho \sigma } ) =
-\partial _{\mu }p - u_{\mu }u^{\rho }\partial _{\rho }p 
\, . 
\label{s}
\end{equation} 

By comparing
(\ref{p}) with (\ref{r}) and (\ref{q}) with (\ref{s}),
we see how the ``kinematical fluid`` has
$\rho \equiv \varepsilon$, zero pressure,
entropy density proportional to the energy one
(i.e. $\sigma \propto \varepsilon$) 
and four-velocity
$n^{\mu }$; thus, on a classical level, our procedure is
equivalent to introduce a real reference fluid behaving as a
non-relativistic dust.\\
Since, being $n_{\mu }\partial _i y^{\mu } = 0$,
the energy-momentum tensor of such a dust is
orthogonal to the hypersurfaces $\Sigma ^3$, then
it contributes
only to the super-Hamiltonian, which rewrites as 
$H^g + H^{\phi } + \sqrt{h} \varepsilon $.

Above, we clarified the physical meaning of the
kinematical action, so
answering for the first posed question, in close analogy
with the approach presented in \cite{KT91} about the
so-called ``Gaussian reference fluid''
(for a discussion on non-Gaussian fluid see \cite{BK95,BK97}).
However, the notion of a reference fluid, here is preserved
also in the (generic) coordinates system $\{ t, x^i\} $,
where, by assigning
the functions $N$ and $N^i$, we get directly the normal field;
in this case, equation (\ref{p}) takes the simple form
$\partial _t{\omega } + \partial _i(N^i{\omega }) = 0$. 
In a synchronous (or Gaussian) reference,
when the fluid is comoving with the coordinates $\{ t, x^i\}$
($N=1\; N ^i=0\; \Rightarrow \;
n^{\bar{\rho }} = (1, {\bf 0})$),
we get, as expected, 
$\omega = \omega (x^i)\; \Rightarrow
\varepsilon = -\omega (x^i)/\sqrt{h}$. 
It is worth noting how, the energy density of
the reference fluid, has the opposite sign of
the super-Hamiltonian
and therefore is, in general, non-positive defined. 

We conclude this section, devoted to the classical aspects of
our reformulation, by stressing the following two points:\\
i) To fix the reference frame, i. e. the lapse function and
the shift vector, via the kinematical action, gives rise
to the appearance of a real dust fluid; by a suggestive
language, we may claim that this ``gauge fixing procedure''
{\em materializes} the reference frame.\\
ii) All the dynamical information about the reference fluid,
result to be contained in the second kinematical equation,
while the first one seems to reflect simply a parameterization
of the dynamics, so ensuring the self-consistency
of the whole theory.

\section{Canonical Quantum Dynamics}

The quantum dynamics, corresponding to the action (\ref{f}), 
is easily got by implementing the Hamiltonian
constraint to their operator form, taken as acting on a wave
functional $\Psi $, now depending even from $y^{\mu }$
(with $\hat{p}_{\mu }=-i\hbar \delta (\; )/\delta y^{\mu }$).
The classical restriction to take a generalized momentum parallel
to the normal field, is translated on a quantum level,
by preserving the super-momentum constraint in its
form and writing down the field equations as 
(the general theory is nevertheless a viable issue)

\begin{equation}   
i\hbar n^{\mu }\frac{\delta \Psi }{\delta y^{\mu }} 
= (\hat{H}^g + \hat{H}^{\phi }_i)\Psi 
\, \quad 
(\hat{H}^g_i + \hat{H}^{\phi }_i)\Psi = 0 
\, \quad \Psi = \Psi (y^{\mu }, \{ h_{ij}\} , \phi ) 
\, . 
\label{t}
\end{equation} 

where the wave functional is taken on the 
3-geometries $\{ h_{ij}\}$, i.e. a whole class of 3-metric
tensors, connected via a 3-diffeomorphism. 

In analogy to the case of the parametrized particle,
we may rewrite this set of $\infty ^3$ equations in the
coordinate system $\{ t, x^i\}$, as soon as, using
the first kinematical equation; 
indeed, as above outlined, 
the use of this (classical) equation is justified,
even on a quantum level, by observing that:\\ 
i) It is necessary to specify the meaning of the field
$n^{\mu }$ in the above quantum dynamics;
indeed, such a field becomes the real normal one 
and the system evolution is constrained
to a fixed choice of $N$ and $N^i$.\\
ii) On a classical level, this equation plies no dynamical
role and resembles very closely an $\infty ^3$-version of
the corresponding equation for the non-relativistic particle.\\
iii) Its use is justified {\em a posteriori} because it reproduces
the expected Schr\"odinger equation.\\
So doing, we get

\begin{equation}   
i\hbar \partial _ty^{\mu }
\frac{\delta \Psi }{\delta y^{\mu }}
= N(\hat{H}^g + \hat{H}^{\phi })\Psi 
\label{u} 
\end{equation} 

By defining the operator
$\partial _t (\; )\equiv \int _{\Sigma ^3}\partial _t y^{\mu }
\delta (\; )/\delta y^{\mu }$ we finally arrive to a single
(smeared) Schr\"odinger equation 

\begin{equation}   
i\hbar \partial _t\Psi = 
i\hbar \int _{\Sigma _t^3} \left\{ 
\frac{\delta \Psi }{\delta y^{\mu }}\partial _ty^{\mu } 
\right\} d^3x =    
\hat{{\cal H}}\Psi  \equiv 
\left[ \int _{\Sigma _t^3} 
N(\hat{H}^g + \hat{H}^{\phi }) d^3x \right] \Psi 
\label{v} 
\end{equation} 

with $\Psi = \Psi (t, \{  h_{ij}\}, \phi )$.
We regard this equation as
the fundamental one of the revised canonical quantum gravity,
which, once fixed $N$ and $N^i$ (the latter one does not
enter in this equation), provides the dynamics of
quantum 3-geometries and matter fields; the label time $t$
acquires a precise physical meaning in view
of the analysis developed in the
previous section, i.e. it is a real ``fluid clock``
filling the hypersurfaces $\Sigma ^3_t$ (the label $t$
specifying each of them); however, to understand the way in which
such a reference fluid manifests its presence on a quantum level,
see the below discussion about the eigenvalues problem.

As shown in \cite{M02}, by adopting a suitable normal
ordering in the kinetic part of $\hat{H}^g$, i.e.
$G_{ijkl}\pi ^{ij}\pi ^{kl}\rightarrow
\delta /\delta h_{ij}(G_{ijkl}\delta /\delta h_{kl})$, then we
are able to turn the space of the solutions into an Hilbert
one by the inner product
$\langle \Psi _1\mid \Psi _2\rangle $
($\Psi _1$ and $\Psi _2$ are generic functionals and 
the bra-ket referring to a functional integral on the space
of all possible 3-geometries). Thus, in such a theory, 
we recognize of 
the evidence for a conserved probability
amplitude $\Psi ^*\Psi $
(being $\Psi^ *$ the complex conjugate of the wave functional),
with $\langle \Psi \mid \Psi \rangle = 1$ and 
$\partial _t \langle \Psi \mid \Psi \rangle = 0$. 

Now we search for the link between the above scheme of
quantization and the notion of reference fluid.
To this end, we expand the wave functional as follows

\begin{equation}  
\Psi (t, \{ h_{ij}\} , \phi ) = 
\int_{ {}^*{\cal Y}_t}D\Omega \Theta (\Omega )
\chi (\Omega , 
\{h_{ij}\} , \phi )
exp\left\{  
-\frac{i}{\hbar }\int _{t_0}^t dt^{\prime }
\int _{\Sigma^3_t}d^3x
(N\Omega )\right\}   
\label{x} 
\end{equation} 

where $D\Omega$ denotes the Lebesgue measure in the
functional space ${}^*y_t$ of the conjugate function  
$\Omega (x^i)$ and $\Theta$ a functional valued in this domain. 
By this form of $\Psi $, equation (\ref{v}) is reduced to the
eigenvalues problem 

\begin{equation}  
(\hat{H}^g + \hat{H^{\phi }})\chi  
= \Omega \chi 
\, .      
\label{y} 
\end{equation} 

Thus we see that, from a quantum point of view,
the label time manifests its physical nature, via the
appearance of a non-zero eigenvalue of the super-Hamiltonian.\\ 
In the limit $\hbar \rightarrow 0$, the WKB
approximation, i.e. 
$\Psi \propto exp\{ i\sigma /\hbar \}$ provides an
Hamilton-Jacobi operator which allows to identify
the two quantities $\omega $  and $\Omega $,
that is to say, on the classical limit, the energy density 
of the (dust) reference fluid is given by
$\varepsilon = -\Omega /\sqrt{h}$;
we will discuss in more detail below
the nature of this identification
(which is valid at all in general),
with respect to the particular
case of the closed FRW cosmology.\\
We see that, since in the coordinates system $\{t, x^i\}$,
the fluid is ``at rest'' to the hypersurfaces $\Sigma ^3_t$,
then it contributes its energy density only to the
super-Hamiltonian eigenvalue (in the limit
$\hbar \rightarrow 0$); Therefore, even from our quantum
analysis of the dynamics, it emerges 

If the theory here proposed is a predictive one,
we should expect to observe 
the trace of this reference fluid
energy density from all those systems which 
underwent a classical limit; such a situation is surely true
for our actual Universe and, indeed, we really observe
(in the synchronous reference of our galaxy)
an unidentified dust energy, the so-called {\em dark matter};
in the next two section, we will try to understand if it
can exist a correlation between our dust fluid and the
observed ``matter component'' of the Universe.

We observe that the restriction
$\partial _iN=0$,
required, on a classical point of view, for the
validity of the dust fluid model, is not so relevant; 
in fact, the real physics is that one observed
by the fluid reference its-self,
(the synchronous one, in the coordinates $\{ t, x^i\}$), as
described by equations (\ref{y})
(which generalize the Wheeler-Dewitt approach). In this sense, 
equation (\ref{v}) provides only a different parameterization
of the real physics. 

To conclude, it is worth remarking how, the main
difference between our approach
and others interesting ones, that lead to the same formal issue
(see the discussion in \cite{M02} about the comparison
with the so-called ``multi-time approach,
as well as the formulations
presented in \cite{R91a,R91b} and \cite{S93,RS93}),
consists of preliminary reducing 
the super-Hamiltonian to a linear form, and, overall, of setting
{\em ad hoc} fields which play the role of time
(for instance in \cite{S93,RS93} is postulated,in the theory, 
the presence of a real mass-less scalar field).
 simply extend to the 3-metric dynamics the
kinematical (embedding-like) action to provide physical
meaning in the splitting procedure, and then interpret it as a
dust fluid (with the role of time). In this scheme the
3-metric is related to the space-time one by
the dynamical field $y^{\mu }$, so, heuristically, we can say to
bypass the {\em theory background independence}.

\section{The Closed FRW Cosmology}

Since the clock by which we are measuring the age of the
Universe is (essentially) a synchronous one, and
we expect the cosmological dynamics became a classical one,
then the contribution of the ``dust fluid'' energy density
must appear in the galaxies recession. Below we will
face the questions about the modifications introduced, by our
approach, in the quantum evolution of the Universe, and 
about the actual value of the dust energy density. 

We investigate the quantum dynamics predicted,
in a synchronous reference, by
equation (\ref{v}) for the closed
Friedmann-Robertson-Walker model
\cite{KT90,H88}, whose line element
reads (below we adopt the standard notations for the
fundamental constants)

\begin{equation} 
ds^2 = -c^2dt^2 + R^2_c(t)[d\xi ^2 + \sin ^2\xi
(d\eta ^2 + \sin ^2\eta d\phi ^2)]
\, \quad
0\le \xi < \pi \;
0\le \eta < \pi \;
0\le \phi < 2\pi \;
\, .
\label{yyy}
\end{equation}

Here $R_c$ denotes the radius of curvature of the Universe,
measurable, in principle, via the relation
$R_c = c/(H\sqrt{ \bar{\Omega } - 1})$
(being $H$ the Hubble function, $\bar{\Omega }$
the critical parameter and
$R_{c(today)}\sim \mathcal{O}(10^{28} cm)$).\\

In the very early phases of the Universe evolution, it is 
expected a space filled by a thermal bath, involving all the
fundamental particles; since, at very high temperatures,
all the massive particles are ultrarelativistic ones, then
the most appropriate phenomenological representation of the
matter-radiation thermal bath, is provided by 
an energy density of the form $\mu ^2 / R_c^4$.\\
Furthermore, the idea that the Universe underwent an inflationary
scenario, leads us to include {\em ab initio} in the dynamics a
real self-interacting scalar field $\phi$,
described by a ``finite-temperature'' potential $V_T(\phi )$
(here $T$ denotes the thermal bath temperature),
which we may take, for instance, in the Coleman-Weinberg form

\begin{equation}
V_T(\phi ) = \frac{B\sigma ^4}{2h^3c^3} +
B\frac{\phi ^4}{hc}\left[ \ln
\left( \frac{l_{Pl}\phi ^2}{\sigma ^2}\right) - \frac{1}{2}\right] + 
\frac{1}{2}{m_T}^2\phi ^2 \, \quad
m_T = \sqrt{\lambda T^2 - m^2} \, \quad
(m, \lambda ) = const. \, ,
\label{CW}
\end{equation}

where $B$ is a parameter related to the fundamental
constraints of the theory (estimated $\mathcal{O}(10^{-3})$,
$\sigma $ corresponds to the
energy scale associated with the symmetry breaking process (i.e.
$\sigma \sim \mathcal{O}(10^{15})GeV)$),
while $m$ and $l_{Pl}$ denote, respectively, the inverse of 
a characteristic lenght and 
$l_{Pl}$ the Planck length
$l_{Pl}\equiv \sqrt{G\hbar /c^3}$.;
the temperature dependence
of the potential term can be also regarded as a time evolution
of the model.\\
The dynamics of such a cosmological model is summarized, 
as shown when developing the Einstein-Hilbert action under
the present symmetries, by the Hamiltonian function

\begin{equation}
\frac{{\cal H}}{c} = -\frac{l_{Pl}^2}{3\pi \hbar }\frac{p_{R_c}^2}{R_c} +
\frac{c}{4\pi ^2}\frac{p_{\phi }^2}{R_c^3} +
\frac{\mu ^2}{R_c} - \frac{3\pi \hbar }{4l_{Pl}^2}R_c +
2\pi ^2R_c^3V_T(\phi )
\, , 
\label{w}
\end{equation}

with $p_{R_c}$ and $p_{\phi }$
being the conjugate momenta to $R_c$ and $\phi$. 

Thus, the Schr\"odinger equation (\ref{v}) reads,
once turned the above Hamiltonian into an operator
(which possesses the right normal ordering), as follows 

\begin{equation} 
\frac{i\hbar }{c}\partial _t\Psi (t, R_c, \phi ) = 
\left\{ \frac{l_{Pl}^2\hbar }{3\pi }
\partial _{R_c}\frac{1}{R_c}\partial _{R_c} - 
\frac{\hbar ^2c}{4\pi ^2}\frac{1}{R_c^3} \partial _{\phi }^2 + 
\frac{\mu ^2}{R_c} - \frac{3\pi \hbar }{4l_{Pl}^2}R_c +
2\pi ^2R_c^3V_T(\phi )\right\} \Psi (t, R_c, \phi ) 
\, , 
\label{z}
\end{equation}

Before going on with the analysis
of this equation, we need to precise some aspects concerning
the potential term relevance during the Universe evolution.\\
It is well-known that the classical scalar field dynamics is
governed by the following equation

\begin{equation}
\ddot{\phi } + 3H\dot{\phi }
+ c^2\hbar ^2\frac{dV_T}{d\phi } = 0
\, . 
\label{cdf}
\end{equation}

The presence of the potential term is surely crucial to generate
the inflationary scenario, but, sufficiently close to
the initial ``Big-Bang'', its dynamical role is expected to be
very limited; in fact, if we neglect the potential term
in (\ref{cdf}), then, remembering that for early times
$R_c\sim \sqrt{t} \; \rightarrow \; H \sim 1/2t$,
we get the free field solution $\phi \propto \ln t$. Now the terms
we retained to solve equation (\ref{cdf}) are potentially
of the order $\mathcal{O}(1/t^2)$; in the limit toward the
``Big-Bang'' ($t\rightarrow 0$), the potential term (\ref{CW})
(we recall that $T\propto 1/R_c\propto `1/\sqrt{t}$)
can be clearly negligible, i.e.
$t^2V_{T(t)}(\phi (t))\; \rightarrow \; 0$.  Apart from
very peculiar stiff cases, all the inflationary potentials
result to be negligible at very high temperatures.

Taking into account the above classical analysis, we may
assume that, during the Planck epoch, when the Universe
performed its
quantum evolution, the potential of the scalar field plies
no significant role; therefore, by 
choosing the following expansion for the wave function 

\begin{equation} 
\Psi (t, R_c, \phi ) =
\int _{-\infty }^{\infty }\int _{-\infty }^{\infty }
d\epsilon dp C(\epsilon ,p)
\theta (\epsilon , p R_c)
exp\{ \frac{i}{\hbar }(p\phi - \epsilon t)\}
\, , 
\label{a1}
\end{equation}

(with $C(\epsilon , p)$ denoting generic coefficients),
we get, from (\ref{z}), the eigenvalues problem

\begin{equation} 
\left\{ \frac{l_{Pl}^2\hbar }{3\pi }
\frac{d\quad }{dR_c}\frac{1}{R_c}\frac{d\quad }{dR_c} + 
\frac{p^2c}{4\pi ^2}\frac{1}{R_c^3} +
\frac{\mu ^2}{R_c} - \frac{3\pi \hbar }{4l_{Pl}^2}R_c
\right\} \theta = \frac{\epsilon }{c}\theta
\, .
\label{a2}
\end{equation}

with the boundary conditions
$\theta (R_c=0)=0$ and $\theta (R_c\rightarrow \infty )=0$. 

A solution to this equation reads in the form

\begin{equation}
\theta \propto \sqrt{R_c}exp\left\{
-\frac{(R_c - R_{c(0)})^2}{4\alpha ^2} \right\} 
\, ; 
\label{a3}
\end{equation}

in order to be the above functional form a solution of equation
(\ref{a2}), we have to require the relations 
$p = \pm \sqrt{\pi \hbar /c}l_{Pl}$,
$\alpha = l_{Pl}/\sqrt{3\pi }$ and 
$\epsilon = -3\pi \hbar cR_{c(0)}/2l_{Pl}^2 $.
Furthermore, since the
ultrarelativistic energy density is manifestly positive,
then, from the following expression for $\mu ^2$

\begin{equation}
\mu ^2 = \frac{l_{Pl}^2}{3\pi }\left(
\frac{1}{2\alpha ^2} - \frac{R_{c(0)}^2}{4\alpha ^4}\right)
\, ; 
\label{mu}
\end{equation} 

we find an important restriction on the continuous
eigenvalue spectrum, i.e.
$-\sqrt{3\pi /2}M_{Pl}c^2 < \epsilon <
\sqrt{3\pi /2}M_{pl}c^2$ (being $M_{pl}$ the Planck mass,
$M_{pl}=\hbar /cl_{Pl}$).\\
Thus, we get a (non-normalizable) probability amplitude, 
for the stationary states, of the form 

\begin{equation}
P_{Stat} \propto \cos ^2(\mid p\mid \phi )
R_cexp\left\{
-\frac{(R_c - R_{c(0)})^2}{2\alpha ^2} \right\} 
\, . 
\label{a4}
\end{equation}

The $\phi $-component of the wave function 
is not normalizable, because of the
potential field absence
(we have to do with a situation analogous to that one of
a free non-relativistic particle admitting only two 
momentum eigenvalues) , but it is remarkable
the existence, as effect of our revised quantization approach, 
of stationary states for the radius of curvature;
in the obtained dynamics, we see that the notion of the
cosmological singularity is replaced by the more physical one 
of a peaked probability to find $R_c$ near zero.
The approximation of neglecting the potential term $V_T$
can be regarded as confirmed {\em a posteriori}
by the small probability that the system penetrates
regions where $R_c$ is much greater than the Planck length and
the temperature is sufficiently small to be compared with the
symmetry breaking scale. 

In order to construct the semiclassical limit of
equation (\ref{a2}), we separate $\theta $ into its
modulus and phase, i.e.
$\theta =\sqrt{\alpha }exp\{ i\beta /\hbar \}$;
then we get the following two, real and complex,
components of equation (\ref{a2}) 

\begin{equation}
-\frac{l_{Pl}^2}{3\pi \hbar }
\frac{1}{R_c}\left( \frac{d\beta }{dR_c}
\right) ^2 +
\frac{p^2c}{4\pi ^2}\frac{1}{R_c^3} + 
\frac{\mu ^2}{R_c} - \frac{3\pi \hbar }{4l_{Pl}^2}R_c
-\frac{\epsilon }{c} 
+ \hbar ^2 V_{Quantum} = 0 \, \quad
\label{a5}
\end{equation} 

\begin{equation}
\frac{1}{\sqrt{\alpha }}
\frac{d\quad }{dR_c}
\left( \frac{\alpha }{R_c}\frac{d\beta }{dR_c}
\right) = 0 \Rightarrow
\alpha \propto R_c/(d\beta /dR_c) 
\, , 
\label{a6}
\end{equation} 

\begin{equation} 
V_{Quantum}\propto
\frac{1}{\sqrt{\alpha }}\frac{d\quad }{dR_c}
\left( \frac{1}{R_c}\frac{d\sqrt{\alpha }}{dR_c}\right) 
\, . 
\label{a7}
\end{equation} 

In the limit $\hbar \rightarrow 0$,
when becomes negligible $V_{Quantum}$, we reobtain the
Hamilton-Jacobi equation describing 
the Universe classical dynamics,
but with an additional term
corresponding to a non-relativistic matter contribution,
which, when $\epsilon $ is negative, acquires
positive energy density; to this respect, we remark how, on the quantum level, 
the Universe is expected to approach the lowest, i.e. negative,
energy state.\\ 
We stress how, for sufficiently large $R_c$, 
if the non-relativistic term dominates
(the spatial curvature being yet negligible),
then we get $d\beta /dR_c \propto \sqrt{R_c}$ and therefore
$R_c\rightarrow \infty \; \Rightarrow 
V_{Quantum}\sim 1/(R_c^3)\rightarrow 0$; such a behavior
supports the idea that, when the Universe ``expands enough''
(i. e. its volume fluctuating explores regions
of high $R_c$ values), it can approach a classical dynamics. 

The analysis of this section answers the question
about the cosmological phenomenology implied by our approach
and the issue goes toward the appearance,
in a synchronous reference, of a
pressureless contribution to the Universe energy density.
In the next section, we make some estimations
in order to understand if such a new term
(which is nothing more than the classical limit of the
total Universe quantum energy) may have something to do with the 
observed dark matter component. 

\section{Phenomenological Considerations}

Indeed, by adding a term to the gravitational action,
we may expect it appears as a new kind of energy-momentum term;
what makes our analysis a valuable one is
in the following points: 

i) The kinematical action is an embedding-like geometrical
object, whose existence in quantum gravity, was postulated
in \cite{M02} on the base of well-grounded statements
and not invented {\em ad hoc}. 
Above we have shown that it can be interpreted,
from a classical point of view, 
as a non-relativistic dust fluid;
a non-relativistic energy density is also what appears
from the quantum dynamics, when taking the classical limit.\\ 
ii) All the accepted models of {\em cold dark matter} predict 
the existence of a very early (decoupled) zero-pressure
component, able, by this feature,
to develop large scale structures
(at the present time even the {\em heat dark matter} is expected to
be non-relativistic). 
Indeed, a non-baryonic component of this kind,
is estimated (either by the supernova data, either
by the cosmic microwaves
background (detected) anisotropy) 
to be $\sim 0.3$ of the actual Universe critical density.\\

Since in equation (\ref{a5}) $\beta$ plies the role 
of the (reduced) action function,
we can write, by using Hamilton equations,
the following relation
\footnote{the same result could be directly obtained
by applying the Hamilton-Jacobi method to the full
action ${\cal S} = \beta (R_c) + p\phi - \epsilon t$.}

\begin{equation}
\frac{d\beta }{dR_c} = p_{R_c} =
-\frac{3\pi \hbar }{2cl_{Pl}^2}
R_c\frac{dR_c}{dt} \, . 
\label{pr}
\end{equation}

Then, remembering that $H = (dR_c/dt)/R_c$ and
$\bar{\Omega } - 1 = c^2/H^2R^2_c$, we see how equation (\ref{a5})
takes the simple form
(with obvious notation for the different contributions) 
$\sum_i X_i = 1$, being 
$X_i\equiv \bar{\Omega }_i/\bar{\Omega }$
($i = p, \mu , dm, curv$); thus, our dust fluid provides
a component of the critical parameter $\bar{\Omega}_{dm}$,
given by

\begin{equation}
\bar{\Omega }_{dm} = -\frac{4l_{Pl}^2c\epsilon }{3\pi \hbar H^2R_c^3}
\, .
\label{odm}
\end{equation}

Such a formula is valid in general,
independently of the other kinds of matter present in the universe,
and, therefore, provides a good tool to investigate the
role it could play in the actual cosmology;
in this respect, we stress the following
three relevant points:\\

i) If we take for $\epsilon$ the minimum value of the
continuous spectrum obtained in the previous
section, within the framework of a ``pre-inflationary''
scenario, i.e.
$\epsilon \sim \mathcal{O}(-M_{Pl}c^2)$, then we get

\begin{equation}
\bar{\Omega }_{dm} = \mathcal{O}\left(
\frac{l_{Pl}c^2H^{-2}}{R_c^3}\right) \sim \mathcal{O}(10^{-63}) 
\, . 
\label{odm1}
\end{equation}

ii) The value of $\epsilon$, required to have 
$\bar{\Omega }_{dm} = \mathcal{O}(1)$
(so that it could make account for the real dark matter
component, estimated about $0.3$ of the
actual critical density), corresponds to

\begin{equation}
\epsilon ^*\sim \mathcal{O}\left(
\frac{\hbar cR_c^3}{l_{Pl}^2c^2H^{-2}}\right) \sim
\mathcal{O}(10^{82}GeV)
\, ; 
\label{odm2}
\end{equation}

such a value corresponds to the present one
of the total energy of the Universe, whether it 
admits a closed space. A crucial point is that
$\epsilon$ is a constant of the motion and therefore,
since the Universe became a classical
one, it was characterized by such value $\epsilon ^*$.\\
iii) In order to get an inflationary scenario, able to
explain the paradoxes of the Standard Cosmological Model,
we need a sufficiently large ``e-folding'' which allows
the size of an horizon, at the inflation beginning,
be now of the order of the actual Hubble radius;
such a value corresponds, at least, to about $60$, i.e.
the ratio between the scale factors, respectively,
after and before the inflation, is around a factor
$\mathcal{O}(10^{26})$. This means that, if today
$R_c\sim \mathcal{O}(10^{28}cm)$, then, taking into account that
the redshift of the end of the inflation is about
$z\sim \mathcal{O}(10^{24})$, we see that when the
de-Sitter phase started its value was
$R_c\sim \mathcal{O}(10^{-22}cm)$.
Thus, the total energy of the Universe, when the
dynamics became dominated by the ``vacuum energy'' at the temperature
$\sigma \sim \mathcal{O}(10^{15}GeV)$, is given by
the expression

\begin{equation}
\epsilon _{\Lambda }\sim
\frac{\sigma ^4R_c^3}{h^3c^3} \sim
\mathcal{O}(10^{36}GeV) \ll \epsilon ^*
\, ;
\label{odm3}
\end{equation}

this result seems to indicate that, assuming
the Universe underwent
an inflationary scenario, we get the contradictory issue
about the impossibility of a dominating
``vacuum energy''.\\
Summarizing, the above considerations are against the idea
that the here obtained $\bar{\Omega }_{dm}$  can make
account for the dark matter, if inflation took place.
The situation is different if we take the picture of the
Standard Cosmological Model because, for instance,
a classical estimation of the thermal bath energy at the
Planck epoch is about
$\mathcal{O}((R_c/l_{Pl})^3M_{Pl}c^2) \sim
\mathcal{O}(10^{112}GeV)$; thus, in absence of inflation,
the value of $\epsilon ^*$ would have become important only
in the later stage of the Universe evolution and it could 
play today a relevant dynamical role.


\begin{thebibliography}{99}

\bibitem{M02}
G. Montani,

{\it Nucl. Phys. B}, (2002), {\bf 634}, 370.

\bibitem{K81} 
K. Kuchar, 
in {\it Quantum Gravity II, a second Oxford symposium}, 
(1981), eds C. J. Isham et al., 
Clarendom Press., Oxford, 

\bibitem{ADM62} 
R. Arnowitt, S. Deser and C, W. Misner, 
in {\it Gravitation: an introduction to current research}, 
(1962), eds I. Witten and J. Wiley, New York. 

\bibitem{D67}
B. S. DeWitt, 
{\it Phys. Rev. }, (1967), {\bf 160}, 1113. 

\bibitem{D97} 
B. S. DeWitt, 
{\it Proc. eighth Marcell Grossmann meeting}, 
(Jerusalem, 22-27 June 1997), ed T. Piran, 6. 

\bibitem{I92} 
C. J. Isham, 
{\it Canonical Quantum Gravity and the Problem of Time}, 
(1992), available /arxiv:/gr-qc/9201011.

\bibitem{R91a} 
C. Rovelli, 
(1991). {\it Class. and Quantum. Grav. }, {\bf 8} (1991) 297. 

\bibitem{R91b} 
C. Rovelli, 
(1991), {\it Phys. Rev. D}, {\bf 43}, 442.

\bibitem{KT91}
K. Kuchar and C. Torre,
{\it Phys. Rev. D}, (1991), {\bf 43}, 419. 

\bibitem{BK95}
J. D. Brown and K. V. Kuchar,
{\it Phys. Rev. D}, {\bf 51}, (1995), 5600.

\bibitem{BK97} 
J. Bick and K. V. Kuchar,
{\it Phys. Rev. D}, {\bf 56}, (1997), 4878.

\bibitem{S93} 
L. Smolin, (1993). available arxiv:gr-qc/9301016. 

\bibitem{RS93}
C. Rovelli and L Smolin, (1993), available arxiv:gr-qc/9308002. 

\bibitem{KT90}
E. W. Kolb and M. S. Turner, 
{\it The Early Universe}, (1990), 
(Adison-Wesley, Reading), 447.  

\bibitem{H88} 
J. B. Hartle, 
in {\it Highlights in Gravitation and Cosmology} (1988), 
eds B. Iver et al., Cambridge Univ. Press. 

\end{thebibliography}
\end{document}